\documentclass[12pt]{iopart}

\usepackage{times}

\usepackage{graphicx}

\usepackage{epsfig,psfrag} 
\usepackage{bbm}
\usepackage{cite}

\newcommand{\omegav}{\omega_{\sf v}}

\begin{document}

\title{Phonon Cavity Models for Quantum Dot Based Qubits}
\author{T. Vorrath$^1$, S. Debald$^1$, B. Kramer$^1$, and T. Brandes$^2$}
\address{$^1$ I. Inst. Theor. Physik, Universit\"at Hamburg, Germany}
\address{$^2$ Department of Physics,
           UMIST,
           P.O. Box 88,
           Manchester 
           M60 1QD, 
           U. K.}

\begin{abstract}
Phonon cavities are believed to be the next step towards a 
control of dephasing in semiconductor quantum dot `qubits'. 
In this paper, we discuss two models for phonon cavities 
-- a surface acoustic wave (SAW)
inter-digitated transducer on an infinite half-space, 
and an elastic thin slab.
The inelastic current  through double quantum dots in 
non-perfect SAW cavities exhibits a gap at small energies
and is completely suppressed in a perfect, infinite system.
In the free-standing slab model, 
van Hove singularities evolve in the phonon spectral density. We find 
that these singularities cause additional side peaks in the inelastic current.
\end{abstract}

\section{Introduction}
Coupled quantum dots exhibit a variety of quantum coherent phenomena 
\cite{Vaartetal95,FujetalTaretal,Blietal98b,Blietal98a,Oosetal98,Qin}
that are expected to be crucial for controlling quantum
superpositions and entanglement. In a semiconductor environment,
these effects depend very sensitively on dephasing by environmental 
bosonic degrees of freedom \cite{BK99,BR00}. The
importance of electron-phonon coupling for transport spectroscopy 
in double quantum dots is well-established by now 
\cite{FujetalTaretal,BK99,Qin}, and suggestions have been 
made \cite{FujetalTaretal,DBK02} to explore 
`semiconductor phonon cavity QED'  in 
nanostructures where phonons and thereby electron-phonon interactions 
become experimentally controllable.

In phonon cavities, sample geometry becomes an important parameter 
since it determines boundary conditions for vibration modes, 
very often in a highly 
non-trivial way  \cite{MM64Aul73} and different from standard 
semiconductor cavity  QED \cite{Yamamoto}.
The control of mechanical properties, shape and boundary conditions
opens the possibility to artificially tune the phonon spectrum of a 
sample. 
In particular, the interaction with phonons of a given energy can be suppressed
which turns out to be important for certain realizations of quantum dot 
based qubits operations \cite{BV02}.

In this paper, we investigate two different ways of modifying 
the phonon modes of
nanostructures based on ideas by Kouwenhoven and van der Wiel \cite{LKWW}.
We discuss the electronic current through two coupled quantum dots 
embedded into phonon cavities.
Double dots act as emitters of phonons \cite{FujetalTaretal,BK99}
and can be used as a measurement device for phonon modes \cite{AK00}, 
where the modifications
in the phonon spectrum can be detected in the inelastic current.

In the first model,
a double dot is placed between two arms of a surface acoustic wave (SAW)
inter-digitated transducer, 
the latter changing the electric boundary conditions
for piezo-electric SAWs in an infinite half-space.
We find that this system forms a
resonator for surface waves leading to a gap in the phonon spectral density,
which should be clearly observable in the current 
through a double quantum dot.

Our second model 
is an idealized version of a {free-standing slab} (thin plate model),
where some confined phonon modes evolve van Hove
singularities while others do not develop any electron-phonon
interaction potential \cite{DBK02}. 
In  a non-perturbative calculation, we find 
that a van Hove singularity leads to additional side peaks in the current.

An important prediction of both models is
the possibility to completely suppress phonon induced dephasing
at phonon frequencies corresponding to certain (tunable) energy 
differences between the dots.

\section{Surface Acoustic Wave Cavity}

\begin{figure}
\centering
\psfrag{0}{$0$}
\psfrag{L}{$l_0$}
\psfrag{x}{$x$}
\psfrag{y}{$y$}
\psfrag{[010]}{[010]}
\psfrag{[100]}{[100]}
\epsfig{file=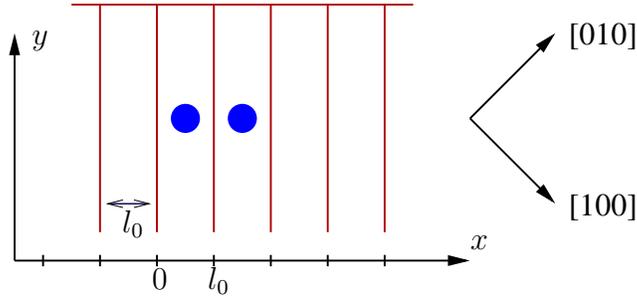,width=8cm}
\caption{Top view on the surface of the sample. The quantum dots (blue)
lie just beneath the surface between the metal stripes (red). 
The $y$-axis is chosen to be parallel to the
stripes and the crystal axes include an angle of 45 degrees with the stripes.}
\label{sample}
\end{figure}

We consider the surface of a semiconductor heterostructure on which thin 
metallic stripes are attached with a constant spacing $l_0$ between each 
other. In an ideal case both the number of stripes and their length is 
assumed to be infinite. Similar structures are used to generate surface
waves by applying an oscillating voltage at resonance frequency between
adjacent stripes. In this work, no voltage between the stripes
is taken into account.
Two coupled quantum dots are located a small distance $z_0$ beneath the
surface at the interface of the heterostructure. Leads are attached to the
quantum dots in order to enable transport experiments. In principle,
all kinds of geometries are feasible as the positions of the quantum dots
as well as the directions of the crystal axes relative to the stripes 
can be chosen at will. However, we restrict our calculation to one certain
geometry where the dots are located symmetrically on both sides of one stripe
exactly in the middle between two stripes as depicted in Fig.~\ref{sample}.
The angle between the stripes and the crystal axes is chosen to be 45 degrees.
The generalization to most other geometries follows directly from that case.

\subsection{Influence on Surface Waves}
Surface acoustic waves propagate along the surface of a medium while their
typical penetration depth into the medium is of the order of one wavelength.
The prediction of their existence dates back to the nineteenth century
and nowadays SAW are used in a wide range of experiments \cite{SAW}.
In piezoelectric media like GaAs the displacement field of the wave generates
an electric potential that dominates the interaction with electrons.
Interaction via the deformation potential is also present but some orders of 
magnitude smaller and is neglected in the following. 
As the piezoelectric potential of the wave has to meet the electric boundary 
conditions at the interface between the medium and the air,
the electron-phonon interaction strongly depends on the electric properties
of the surface.

In our model any mechanical influence of the metallic stripes on the elastic
properties of the medium is neglected. Instead, we expect an additional 
boundary condition due to the metallic stripes, as they are all connected 
to each other and form an equipotential line on the surface,
\begin{equation}
\label{condition}
\varphi(x\!=\!n l_0, y) = \mbox{const.}, \qquad n\in \mathbbm{Z}.
\end{equation}
The $x$ and $y$ coordinates are chosen such that the stripes are parallel 
to the $y$-axis and located at positions $x\!=\!nl_0$ while
the width of the stripes is neglected.
We assume that all surface modes, whose piezoelectric potential does not 
fulfill the boundary condition (\ref{condition}) are suppressed and 
consequently need not to be considered any further. Thus, the remaining
task is to find all surface waves compatible with (\ref{condition}).

Without additional boundary conditions as (\ref{condition}), surface waves
propagate as plane waves with wave vector $\mathbf{q}\!=\!(q_x,q_y)$ along 
the surface.
As the wave equation for the displacement field is linear, any linear 
combination of plane surface waves again gives a solution of the wave equation.
Hence, we consider the antisymmetric combination $\mathbf{w}_\mathbf{q}$ 
of surface plane waves with
wave vectors $(q_x,q_y)$ and $(-q_x,q_y)$ whose displacement field is given by
\begin{equation}
\label{displacement}
\mathbf{w_q}(\mathbf{r},t) = C e^{i(q_y y -\omega t)} \left(
\begin{array}{c}
  a(q,z) \cos(\alpha) \cos(q_x x) \\
  i a(q,z) \sin(\alpha) \sin(q_x x) \\
  -b(q,z) \sin(q_x x)
\end{array} \right).
\end{equation}
In an isotropic model of the crystal, the functions $a(q,z)$ and $b(q,z)$
describe the decay of the SAW amplitude in the depth $z$ of the medium and 
$\alpha$ is the angle between the $x$-axis and the wave vector $\mathbf{q}$.
This combination corresponds to a standing wave in $x$-direction and a 
plane wave in $y$-direction. For the piezoelectric potential created by this
mode we find
\begin{equation}
\label{piezo}
\varphi(\mathbf{r},t) = - C \frac{e_{14}}{\varepsilon_0 \varepsilon} \;
    \big(\!\cos^2(\alpha) - \sin^2(\alpha)\big)\, 
    f(qz) \sin(q_x x) \, e^{i(q_y y -\omega t)},
\end{equation}
with $e_{14}$ the piezoelectric stress constant, $\varepsilon_0$ the 
dielectric constant, $\varepsilon$ the relative permitivity of the medium.
The function $f(qz)$ describes the decay in $z$-direction \cite{ICPS} 
and follows from the boundary condition for the electric field on the surface.
Here, we assume a free surface, i.e. a non-conducting surface, apart from
condition (\ref{condition}). Due to the special form of the 
mode $\mathbf{w}_\mathbf{q}$ as a standing wave in $x$-direction, 
the corresponding piezoelectric potential (\ref{piezo}) vanishes at the
lines $x\!=\!m\pi/q_x$. Thus, the additional boundary condition 
(\ref{condition}) is met by the modes $\mathbf{w}_\mathbf{q}$ if and only if
\begin{equation}
\label{qx}
q_x = m \, \frac{\pi}{l_0}, \qquad m \in \mathbbm{N}.
\end{equation}
Apart from the plane surface waves with $q_x\!=\!m 2\pi/l_0$ and $q_y\!=\!0$,
no other modes are compatible with condition (\ref{condition}).

\subsection{Electron-SAW Interaction}
We consider interactions between electrons and surface waves due to the 
piezoelectric potential (\ref{piezo}). 
For the calculation of the inelastic current between two quantum dots,
it is necessary to find the quantized form of the surface modes 
$\mathbf{w}_\mathbf{q}$ and the corresponding piezoelectric potential.
This can be achieved by calculating the Lagrange function for these modes
in a quantization volume of area $L \cdot \!L$ and infinite depth.
Displacement and momentum can be replaced by operators obeying the 
canonical commutation relations. Finally the Hamiltonian of the surface 
waves can be written as a sum of harmonic oscillators, i.e. surface phonons. 
This specifies the normalization constant $C$ in (\ref{displacement}) 
and (\ref{piezo}) as
\begin{equation}
C = \frac{1}{L} \sqrt{\frac{\hbar}{\rho \lambda v}}, 
\end{equation}
where $\rho$ is the density of the medium, $\lambda$ a material parameter
and $v$ the velocity of the SAW. Note, that $C$ does not depend on the 
wave vector $\mathbf{q}$ but on the quantization area $L^2$.
Thus, the electron-phonon interaction Hamiltonian writes
\begin{equation}
\label{int}
H_{\sf int} = \sum_{\mathbf{q}} \gamma_{\mathbf{q}}(\mathbf{r})\;
   \Big( b_{q_x, q_y}+ b_{q_x,-q_y}^{\dagger} \Big).
\end{equation}
The boson operator $b_{\mathbf{q}}^{\dagger}$ creates a surface phonon
with the displacement field $\mathbf{w}_\mathbf{q}(\mathbf{r},t)$
as given in (\ref{displacement}). The interaction coefficient 
$\gamma_{\mathbf{q}}$ is defined as
\begin{equation}
\gamma_{\mathbf{q}}(\mathbf{r}) = 
   C \; \frac{e e_{14}}{\varepsilon_0 \varepsilon} \; 
   \big(\!\cos^2(\alpha) - \sin^2(\alpha)\big)\, 
   f(qz) \sin(q_x x) \, e^{i q_y y} \, .
\end{equation}

\subsection{Inelastic Current through a Double Quantum Dot}
We calculate an approximation for the inelastic current through a 
double quantum dot for the geometry described above, when only 
interactions with phonons $\mathbf{w}_\mathbf{q}$ (\ref{displacement})
with (\ref{qx}) are considered.
A source-drain voltage between the leads on both sides of the double 
quantum dot gives a current through the double quantum dot, as electrons 
tunnel step by step from one lead to one dot, then to the other dot and finally
to the other lead. In the Coulomb blockade regime, only one additional 
electron is on either of the dots at the same time.
If, however, a bias voltage is applied between the two quantum dots, the
mismatch of the energy levels of the additional electron in the two dots
results in a suppression of the current due to energy conservation.
In that case, only inelastic processes like emission or absorption of 
phonons allow a finite current \cite{FujetalTaretal}.
At zero temperature this inelastic current can be approximated by
\begin{equation}
\label{I_inel}
I_{\sf inel} = 2\pi \, T_c^2 \;\rho(\varepsilon), 
\end{equation}
where $\varepsilon$ is the energy difference of the additional electron
between the two quantum dots \cite{BK99}. The tunneling rate between the dots 
is given by $T_c$ and the effective spectral density is defined as
\begin{equation}
\label{rho}
\rho(\omega) = \sum_{\mathbf{q}} 
\frac{|\alpha_{\mathbf{q}}-\beta_\mathbf{q}|^2}{\hbar^2 \omega^2} \;
\delta(\omega - \omega_\mathbf{q}).
\end{equation}
Here, $\alpha_{\mathbf{q}}$ and $\beta_\mathbf{q}$ are the matrix
elements of the interaction operator (\ref{int}) with the electron wave 
function in the left or the right quantum dot, respectively.
The size of the quantum dot is assumed to be negligible against the 
phonon wavelength and hence the electron density is approximated by a 
$\delta$ function at the center of the corresponding quantum dot.
Taking into account only wave numbers with $q_x$ as given in (\ref{qx})
we find
\begin{equation}
|\alpha_{\mathbf{q}}-\beta_\mathbf{q}|^2 = 
\left\{ \begin{array}{l l}
 4 \gamma_\mathbf{q}^2 (z_0) \quad & m \mbox{ odd} \\[1mm]
 0 &m \mbox{ even}
\end{array} \right.
\end{equation}
The modes with even $m$ do not contribute to the current because the
corresponding piezoelectric potential has zeros at the positions of the
quantum dots and therefore these modes do not interact with the dots.

In the thermodynamic limit of a macroscopic sample, the $y$-component of the
SAW wavenumber becomes quasi continuous and the sum over $q_y$ can be 
replaced by an integral. Thereby the system-length $L$ cancels and the
effective spectral density writes
\begin{equation}
\hspace*{-15mm}
\rho(\omega) = \frac{2}{\pi \hbar \omega^2 \rho \lambda v^2}
\left( \frac{e e_{14}}{\varepsilon_0 \varepsilon} \right)^2
\frac{1}{L} \sum_{q_x} \int \! dq_y \, \delta\Big(\frac{\omega}{v}-q\Big)
\big(\!\cos^2(\alpha) - \sin^2(\alpha)\big)^2\, f^2(q z_0).
\end{equation}
The same procedure does not apply for the sum over $q_x$ since these modes
do not depend on the system-length $L$ but obey condition (\ref{qx}).
Hence, the inelastic current through the double quantum dots still depends 
on the system-length $L$ and will vanish
in the thermodynamic limit of $L\!\to\!\infty$.
The reason is that the energy $\hbar \omega_q$ of one phonon is distributed
over the whole sample. By increasing the system-size $L$, the amplitude 
of the displacement and the piezoelectric potential, and therewith
the interaction strength is decreased and finally vanishes.
Normally, this effect is canceled by an increasing number of modes within each
interval of energy. Here, however, the modes are independent of the 
system-size due to the additional boundary condition made by the stripes.
Thus, the inelastic current through the double quantum dots
due to emission of surface waves
is completely suppressed by the metallic stripes in the ideal case.
\begin{figure}
\centering
\psfrag{x}{\hspace*{-3mm}$\omega / \omega_0$}
\psfrag{rho}{\hspace*{-6mm}$\rho(\omega) N \omega_0$}
\epsfig{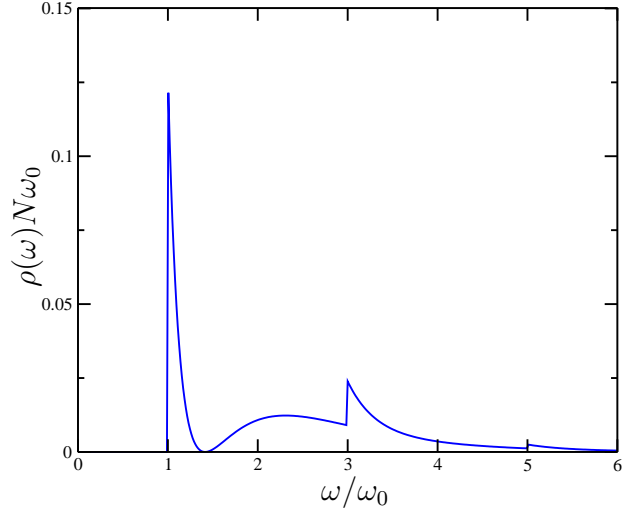}
\caption{Inelastic current through a double quantum dot
with finite system-length~$L\!=\!Nl_0$. The quantum dots are $z_0\!=\!100$
nm beneath the surface and their distance is equal to the spacing of
the metallic stripes $l_0\!=\!250$ nm. Material parameters are taken for GaAs.
The frequency $\omega_0$ corresponds to an energy of 22 $\mu$eV.}
\label{step}
\end{figure}

In a real sample with disorder and impurities, the phonon is not extended
over the whole sample size but concentrated on a finite area, leading to
a non-vanishing inelastic current. Moreover, the suppression of phonon modes
with wavenumber $q_x$ close to condition (\ref{qx}) is not perfect.
These modes are damped but still interact with the quantum dots and hence
ensure a finite inelastic current. The width of the interval 
$\Delta q_x$ around $q_x\!=\!m\pi/l_0$ in which the suppression of phonon
modes is not complete depends on the quality of the SAW cavity formed by
the stripes on top of the sample.
We include both effects in a finite system-length $L$ and find 
\begin{equation}
\label{current}
\hspace*{-15mm}
\rho(\omega) N \omega_0 =
 \frac{4}{\pi^2 \hbar  \rho \lambda v^3}
\left( \frac{e e_{14}}{\varepsilon_0 \varepsilon} \right)^2
\left( \frac{\omega_0}{\omega} \right)^2 f^2(\omega z_0/v) 
\sum_{m=1,3,...}^{m<\omega/\omega_0}
\Big[ 2 m^2 \left( \frac{\omega_0}{\omega} \right)^2 -1\Big]^2,
\end{equation}
where the system-length is given in units of the spacing of the stripes,
$L\!=\!Nl_0$ and a typical frequency for the system is introduced,
$\omega_0\!=\!\pi v/l_0 $. This approximation for the inelastic current 
is shown for typical experimental values in Fig.~\ref{step}.

Summarizing one can say that the metallic stripes on top of the sample 
form a cavity for surface acoustic waves since the additional boundary 
condition require a standing wave in one direction. 
As a consequence, the inelastic current caused by emission of surface 
waves differs extremely from the case without stripes \cite{ICPS}.
In the ideal case, the inelastic current is completely suppressed.
However, taking into account a non-perfect cavity due to a finite
length, a finite width, and a finite number of stripes we find a non-vanishing
inelastic current as given in Eq.~(\ref{current}).

For energies smaller than $\hbar \omega_0$ the lowest standing wave mode can
not be excited and consequently the inelastic current exhibits a gap in that
energy region. The excitation of higher modes manifests itself in a step in
the inelastic current, as can be seen in Fig.~\ref{step} for 
$\omega/\omega_0\!=\!1,3,5,\ldots$ . Furthermore, at 
$\omega/\omega_0\!=\!\sqrt{2}$ the inelastic current vanishes, because 
the SAW would be emitted along the crystal axes in this case, but
there is no piezoelectric interaction in that direction.
It seems promising to investigate the influence of a SAW cavity also for
other geometries. If the quantum dots are realized right beneath the 
metallic stripes, the inelastic current will be further suppressed.
Moreover, the angle between the stripes and the dots or between the 
stripes and the crystal axes can be changed.

\section{Phonon Cavities}
Another possibility to change the phonon spectrum of a sample is to
use geometrical confinement and thereby changing the
mechanical boundary conditions of the system.

Recently, considerable progress has been made in the fabrication of
nano-structures (`phonon cavities') that are only partly suspended or
even free-standing \cite{CR98,BRWB98}. They considerably differ in
their mechanical properties from the bulk material. For example,
phonon modes are split into several subbands, and quantization effects
become important for the thermal conductivity \cite{SW92,GRKV97,RK98}.

Moreover, we have recently predicted that phonon subband
quantization can also be detected in the non-linear electron current
through double quantum dots embedded into nano-size semiconductor
slabs acting as phonon cavities \cite{DBK02}. Like in the previous
section, the phonons act as a source of dissipation for the electrons
by spontaneous emission even at zero temperature. These inelastic
processes lead to fingerprints of the phonon density of states in the
electron current (see Eqs.~(\ref{I_inel}) and (\ref{rho})) showing
quantitatively the possibility to use double quantum dots as phonon
detectors \cite{AK00}. On the other hand, double dots can be used as
tunable energy selective phonon emitter \cite{FujetalTaretal} with
well defined emission characteristics because the transport is
mediated by spontaneous emission.

\begin{figure}[t]
\begin{center}
\epsfig{file=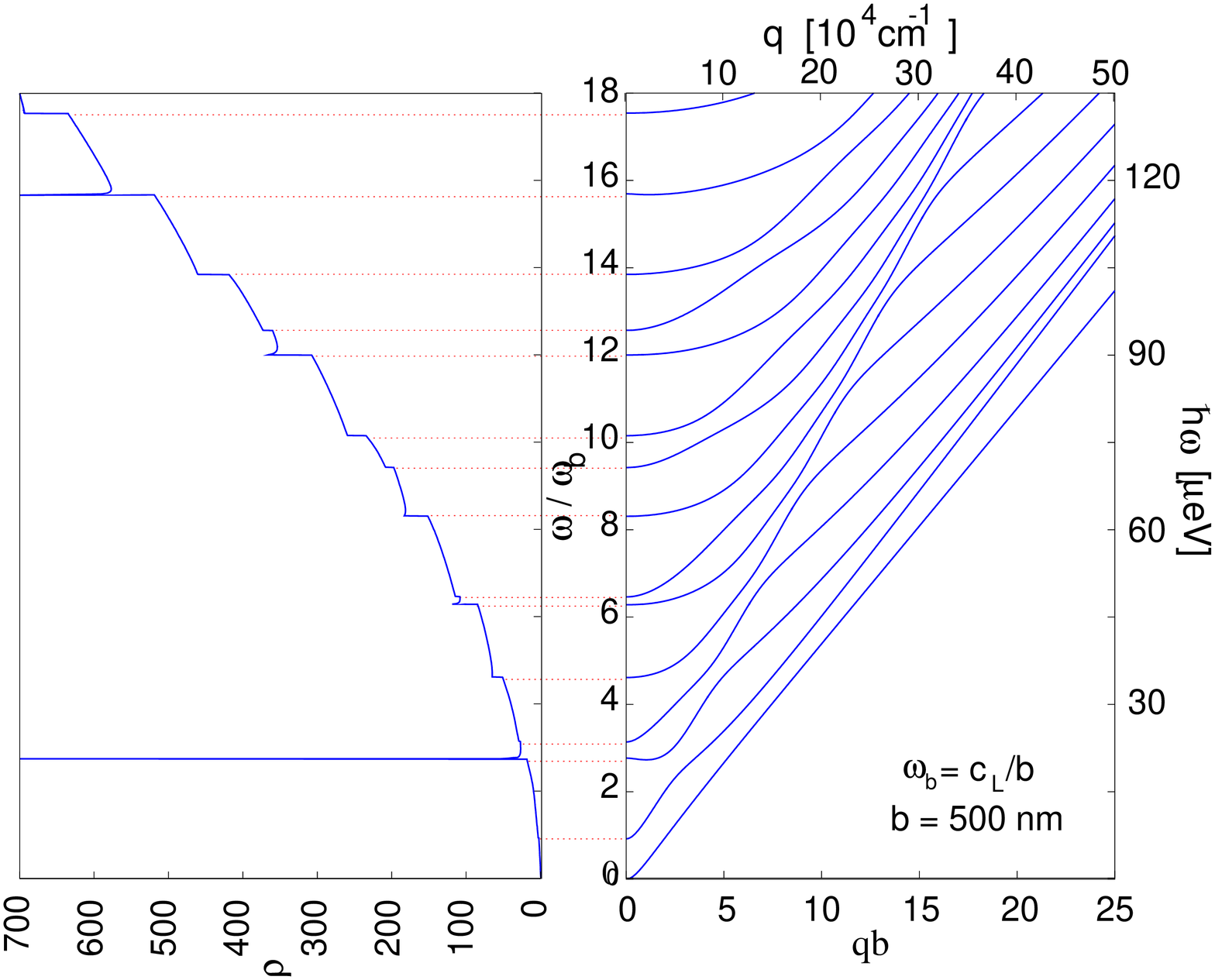, width=10.4cm}
\end{center}
\caption{Dispersion relation (right) and density of states (left) of
typical cavity phonons. The characteristic energy is given by $\hbar \omega_b =
\hbar c_l / b$ with the longitudinal speed of sound $c_l$ and the cavity width
$2b$. For a GaAs planar cavity of width $2b=1\,\mu$m the energy is given by
$\hbar \omega_b=7.5\,\mu$eV. The minimum in the dispersion of the third subband
leads to a van Hove singularity in the phonon DOS at $\hbar \omegav \approx
2.5 \, \hbar \omega_b$.} 
\label{fig-dos}
\end{figure}

This fact together with some peculiar properties of phonons confined
to nanosize planar cavities open the path to new and interesting
effects.  For instance, at a certain energy $ \hbar \omega_0 $ the
cavity phonon corresponding to the frequency $ \omega_0 $ evolves a
displacement field $\mathbf{w}(\mathbf{r},t)$ of the cavity that does
not induce any interaction potential (piezoelectric \textit{or}
deformation potential) acting on the electrons \cite{DBK02}. This
corresponds to a {\em complete decoupling} of dot electrons and cavity
phonons leading to the possibility to suppress phonon induced
dephasing in double quantum dot qubits. A further peculiarity of
cavity phonons is that certain subbands lead to the existence of van
Hove singularities in the density of states. These correspond to
minima in the dispersion at finite wavevectors with preceeding
negative phonon group velocity, see Fig.~\ref{fig-dos}. \, A van Hove
singularity leads to strong inelastic transitions in the double
quantum dot corresponding to a large emission rate of phonons of
energy $\hbar \omegav$.

In the following we investigate the influence of such van Hove
singularities on the electron transport through coupled quantum dots
in a non-perturbative way.  Led by previous numerical results
\cite{DBK02} we decompose the phonon spectral density (\ref{rho}) to 
find an analytical model
\begin{equation}
\rho(\omega) = \rho_{\sf Ohm} + \rho_0 \delta(\omega - \omegav)
\end{equation}
consisting of a van Hove singularity at energy $\hbar \omegav$ and a
background of lower order subbands that form an ohmic bath.  We follow
previous calculations \cite{BK99} using a combination of a master
equation approach and a unitary transformation. Neglecting the ohmic
bath, the stationary inelastic current at zero temperature can be
written as
 \begin{equation}
 I_{\sf inel} \approx 2 \pi T_c^2
\sum_{n}\gamma_n \delta(\omega - n\omegav) .
\end{equation}
Thus, besides the main peak at $\hbar \omegav$, the van Hove
singularity leads to non-perturbative satellite peaks at harmonics of
the frequency $\omegav$ with oscillator strength $\gamma_n$ similar to
double dot systems under monochromatic microwave irradiation \cite{SN96}.
The influence of the ohmic bath background manifests in a
power law divergence \cite{BK99} instead of the former deltalike
singularities at energies $n\hbar\omegav$.

\section{Conclusions}
Designing the geometry of nanostructures allows to create 
phonon cavities with tailored  phonon spectra
and electron-phonon interaction. The latter can be 
completely suppressed at certain energies (or even energy windows)
which is expected to be crucial for the realization of 
semiconductor quantum dot based quantum information processing.
We have
shown that emission of surface acoustic phonons can be 
completely suppressed in coupled quantum dots embedded between
an infinite inter-digitated transducer. Moreover,
in a free-standing slab, double quantum dots can be brought into resonance with
zeros of the phonon spectral density (leading to a vanishing dephasing in 
lowest order), or
with van Hove singularities of the phonon spectrum which lead to 
additional side peaks in the inelastic current.

This work was supported by the EU via TMR and RTN projects
FMRX-CT98-0180 and HPRN-CT2000-0144, DFG projects Kr~627/9-1, Br~1528/4-1, 
the UK project EPSRC~R44690/01, and the 
UK Quantum Circuits Network. Discussions with R. H. Blick,
T.~Fujisawa, W. G. van der Wiel, and L. P. Kouwenhoven are gratefully
acknowledged.

\end{document}